\documentclass[aps,showpacs,eqsecnum,twocolumn,superscriptaddress]{revtex4}

\usepackage{amsmath,amssymb,latexsym}
\usepackage{graphicx,xcolor}

\newcommand{\be}{\begin{equation}}
\newcommand{\ee}{\end{equation}}
\newcommand{\bea}{\begin{eqnarray}}
\newcommand{\eea}{\end{eqnarray}}
\newcommand{\gm}{\gamma}
\newcommand{\tgm}{\tilde{\gamma}}

\begin{document}

\title{Excision scheme for black holes in constrained evolution 
formulations: Spherically symmetric case}

\author{Isabel Cordero-Carri\'{o}n}
\affiliation{Laboratoire Univers et Th\'eories (LUTH), Observatoire de
Paris/CNRS/Universit\'e Paris Diderot, 5 place Jules Janssen, F-92190 Meudon, 
France}
\affiliation{Max-Planck-Institute f{\"u}r Astrophysik, 
Karl-Schwarzschild-Str. 1, D-85748, Garching bei M{\"u}nchen, Germany}

\author{Nicolas Vasset}
\affiliation{Laboratoire Univers et Th\'eories (LUTH), Observatoire de
Paris/CNRS/Universit\'e Paris Diderot, 5 place Jules Janssen, F-92190 Meudon, 
France}

\author{J\'er\^ome Novak}
\affiliation{Laboratoire Univers et Th\'eories (LUTH), Observatoire de
Paris/CNRS/Universit\'e Paris Diderot, 5 place Jules Janssen, F-92190 Meudon, 
France}

\author{Jos\'e Luis Jaramillo}
\affiliation{Max Planck Institute for Gravitational Physics (Albert Einstein 
Institute), Am M\"uhlenberg 1, D-14476 Golm, Germany}

\date{\today}

\begin{abstract}
  Excision techniques are used in order to deal with black holes in
  numerical simulations of Einstein's equations and consist in removing
  a topological sphere containing the physical singularity from the
  numerical domain, applying instead appropriate boundary conditions
  at the excised surface. In this work we present recent developments
  of this technique in the case of constrained formulations of
  Einstein's equations and for spherically symmetric spacetimes. We
  present a new set of boundary conditions to apply to the elliptic
  system in the fully constrained formalism of Bonazzola \textit{et al.}
	[Phys. Rev. D 70, 104007 (2004)], at an arbitrary coordinate sphere inside
  the apparent horizon. Analytical properties of this system of
  boundary conditions are studied and, under some assumptions, an
  exponential convergence toward the stationary solution is exhibited
  for the vacuum spacetime. This is verified in numerical examples,
  together with the applicability in the case of the accretion of a
  scalar field onto a Schwarzschild black hole. We also present the
  successful use of the excision technique in the collapse of a
  neutron star to a black hole, when excision is switched on during
  the simulation, after the formation of the apparent horizon.
  This allows the accretion of matter remaining outside the excision
  surface and for the stable long-term evolution of the newly formed
  black hole.
\end{abstract}


\maketitle

\section{Introduction}
\label{sect:introduction}
Relativistic simulations of astrophysical phenomena involving one or several 
black holes (BHs) have undergone significant improvements in the last decade, 
in particular with the first successful studies of binary BH 
systems~\cite{pretorius-05, campanelli-06, baker-06}. One of the major 
difficulties in performing such simulations is the handling of the physical 
singularity of the BH, where some physical fields may diverge. In order to cope
with this problem, essentially two types of methods have been proposed in the 
literature: i)~\emph{excision}, where the singularity, together with its 
neighborhood, is removed from the computational domain and eventually replaced 
by boundary conditions (see e.g. Refs.~\cite{seidel-92, lehner-00}); and 
ii)~\emph{punctures}, where the BH is set in such initial data that the 
physical singularity is not included, but instead the spatial hypersurface 
containing the initial data follows a wormhole through to another copy of 
spacetime, which is compactified and its infinity is reduced to a point, the 
``puncture'' (see e.g. Refs.~\cite{BraBru97,hannam-07}). The wormhole topology is 
prescribed analytically in the conformal factor [see Eq.~(\ref{e:conf_metric}) 
below], which diverges at the puncture location.

Both of these approaches have been successfully applied in simulations of 
binary BH systems, but with different formulations of Einstein equations and 
gauge choices: excision has been used in conjunction with the generalized 
harmonic gauge~\cite{pretorius-05, boyle-07}, and punctures have usually been 
associated with the so-called BSSN (from Baumgarte-Shapiro~\cite{baumgarte-99} 
Shibata-Nakamura~\cite{shibata-95}) formulation~\cite{campanelli-06,baker-06}. 
All these studies use free evolution schemes, in which the constraint equations
arising in the 3+1 decomposition of Einstein equations are not enforced during 
the evolution. If the constraint equations are satisfied initially, they are 
also satisfied during the evolution theoretically, but this is not necessarily 
the case in numerical simulations. In these formulations, the constraint 
equations are satisfied by the initial data and then monitored during the 
evolution to check the validity of the numerical solution.

In such free evolution schemes, most of the resulting partial differential 
equations (PDEs), coming from Einstein and matter equations (in the case of 
nonvacuum spacetimes), are of hyperbolic type. In particular, when suitable 
gauge conditions are chosen, their characteristics, computed inside the BH 
(apparent) horizon, are all directed towards where the singularity is located 
in the spacelike hypersurface. This means that in these schemes, within the 
excision approach and adopting excision surfaces lying inside the apparent 
horizon (AH) such that their evolution world tubes are of spacelike character, 
there is no need for imposing any inner boundary condition (see, e.g., 
Ref.~\cite{Dorband06}, where the excision sphere was placed sufficiently close 
to the horizon).

However, when solving constraints arising in the 3+1 formulation of Einstein's 
equations, the elliptic nature of these PDEs requires that correct 
corresponding boundary conditions at the excision surface have to be defined 
and tested. Otherwise, incorrect boundary conditions will not give the correct 
physical content for the numerical solution and, therefore invalidate the whole 
simulation. This is particularly true in the case of a Fully Constrained 
Formulation~(FCF), such as the one devised by Bonazzola \textit{et al.}~\cite{FCF}, 
where the constraint equations are regularly solved as part of the elliptic set
of equations and enforced during the numerical simulation. It has been checked 
that in the FCF all the characteristics of the hyperbolic sector point towards 
the singularity when a coordinate system adapted to the dynamical spacelike 
excision world tube is used in the evolution \cite{Jaramillo}.

A geometric approach to defining proper boundary conditions for the elliptic part
of Einstein's equations has been undertaken on the basis of the isolated horizon 
paradigm~\cite{CooPfe04, jaramillo-04, DaiJarKri04, gourgoulhon-06, 
CauCooGri06, Jaramillo:2009cc} or the dynamical ``trapping horizon'' 
concept~\cite{Jaramillo}. They have been successfully applied to stationary 
spacetimes~\cite{jaramillo-07, vasset-09}, which can then be used as initial 
data for further dynamical evolutions.

In this paper, we propose a different approach for the definition of boundary 
conditions in the FCF formulation using the excision technique in dynamical 
spacetimes, with the motivation of simulating astrophysical scenarios like a 
star collapsing to a BH. In this context, initial data are usually regular and 
no BH is present yet. During the simulation, a BH forms and, particularly when 
using singularity-avoiding time coordinates (e.g. maximal slicing), an AH is 
found before the appearance of the physical singularity~\cite{shapiro-80}. When
no symmetry is assumed, the AH does not have a simple shape in general and, 
therefore, it is not easy to use it numerically as an excision surface to 
impose boundary conditions. Here, we suggest using an arbitrary but nearby 
sphere inside the AH to define simple and appropriate boundary conditions for 
the elliptic system of PDEs in the FCF of Einstein's equations, in the particular
case of spherical symmetry. Numerical codes assuming spherical symmetry which 
include complex microphysics are nowadays still relevant; for example, 
spherically symmetric relativistic simulations were used to study the influence
of different microphysics and equations of state in the formation of the BH in 
Refs.~\cite{Peres13, Nakazato12, OConnor11}, instead of using simulations without any
symmetry assumptions, due to the very long simulation times involved. In the 
case of no symmetry assumptions, we plan to follow similar ideas for using the 
excision technique in dynamical evolutions, but this study is beyond the scope 
of the present work.

We will describe how the excision technique is used in dynamical evolutions and
show the practical applicability of this approach in the case of a 
Schwarzschild BH spacetime, the accretion of a scalar field into a spherical BH
and the collapse of a neutron star to a BH. Previous works by Scheel 
\textit{et al.}~\cite{scheel-95} and Rinne and Moncrief~\cite{rinne-13} have 
presented similar approaches in constrained formulations in spherical symmetry,
too. Scheel \textit{et al.}~\cite{scheel-95} have considered the situation of 
dust collapse in the Brans-Dicke theory of gravity, whereas Rinne and
Moncrief~\cite{rinne-13} have studied scalar and Yang-Mills fields coupled to 
gravity in a constant-mean-curvature slicing. Scheel 
\textit{et al.}~\cite{scheel-95} set the excision boundary at the AH, at a 
fixed radial coordinate;  the main difference with this work is that we here 
set the excision boundary to be an arbitrary sphere located strictly inside the
AH, and let this AH evolve in time. This approach allows in particular for a 
very straightforward extension to spacetimes without symmetries, where the AH 
can form with a shape deviating from a coordinate sphere. The latter approach 
was also followed in the work of Rinne and Moncrief~\cite{rinne-13}, where the 
value of the conformal lapse function was frozen and evolution equations were 
used to update the values of the remaining variables at the excision surface. 
Here we propose and analyze a different prescription for the boundary 
conditions at the excision surface, that underline the geometric features of 
the system.

The paper is organized as follows. The FCF of Einstein's equations is briefly 
overviewed in Sec.~II, which also contains some remarks in the spherically 
symmetric case. In Sec.~III we describe the excision method, including our 
excision region and the boundary conditions imposed on each slice. Section IV 
discusses numerical results. A summary of our conclusions is given in Sec.~V. 
We use units in which $c = G = M_{\odot} = 1$. Greek indices run from 0 to 3, 
latin indices from 1 to 3, and we adopt the standard convention for the 
summation over repeated indices. $\partial_\alpha$ denotes partial derivatives.


\section{Fully Constrained Formulation}
\label{sect:FCF}
Given an asymptotically flat spacetime $({\cal M}, g_{\mu\nu})$, we consider a 
3+1 splitting by spacelike hypersurfaces $\Sigma_t$, denoting timelike unit 
normals to $\Sigma_t$ by $n^\mu$. The spacetime on each spacelike hypersurface 
$\Sigma_t$ is described by the pair $(\gm_{ij},K^{ij})$, where 
$\gm_{\mu\nu} = g_{\mu\nu} + n_\mu n_\nu$ is the Riemannian metric induced on 
$\Sigma_t$. We choose the convention 
$K_{\mu\nu} = - \frac{1}{2} {\cal L}_{\bf{n}} \gm_{\mu\nu}$ for the extrinsic 
curvature. With the lapse function $N$ and the shift vector $\beta^i$, the 
Lorentzian metric $g_{\mu\nu}$ in the 3+1 formalism can be expressed in 
coordinates $(x^\mu)$ as
\be
	g_{\mu\nu} dx^\mu dx^\nu = - N^2 dt^2 + \gm_{ij} (dx^i + \beta^i dt) 
(dx^j + \beta^j dt).
\ee

As in Ref.~\cite{FCF}, we introduce a time-independent flat metric $f_{ij}$, which 
satisfies ${\cal L}_{\bf{t}} f_{ij} = \partial_t f_{ij} = 0$ and coincides with
$\gm_{ij}$ at spatial infinity. With the definitions $\gm := \det \gm_{ij}$ and
$f := \det f_{ij}$, we introduce the following conformal decomposition of the 
spatial metric,
\be
	\gm_{ij} = \psi^4 \tgm_{ij}, \;\;\; \psi = (\gm / f)^{1/12}.
\label{e:conf_metric}
\ee
The difference between the conformal metric and the flat fiducial one is 
denoted by $h^{ij}$, $h^{ij} := \tgm^{ij} - f^{ij}$. The chosen prescriptions 
for the gauge variables in Ref.~\cite{FCF} are the maximal slicing,
\be
	K = 0,
\ee
and the so-called generalized Dirac gauge,
\be
	{\cal D}_i \tgm^{ij} = {\cal D}_i h^{ij} = 0,
\label{e:Dirac}
\ee
where $K = \gm^{ij} K_{ij}$ denotes the trace of the extrinsic curvature and 
${\cal D}_k$ stands for the Levi-Civita connection associated with the flat 
metric $f_{ij}$. Finally, we introduce the conformal decomposition
\be
	\hat{A}^{ij} := \psi^{10} K^{ij}.
\ee
In this formulation, Einstein's equations result in a coupled elliptic-hyperbolic
system: the elliptic sector acts on the variables $\psi$, $N$ and $\beta^i$, 
while the hyperbolic sector acts on $h^{ij}$ and $\hat{A}^{ij}$. More details 
of the analysis of both sectors can be found in Refs.~\cite{CC08, CC09}.

The decomposition of $\hat{A}^{ij}$ in longitudinal and transverse-traceless 
(TT) parts 
\be
	\hat{A}^{ij} = (LX)^{ij} + \hat{A}^{ij}_{\mathrm{TT}},
\ee
where $(LX)^{ij} := {\cal D}^i X^j + {\cal D}^j X^i - \frac{2}{3} f^{ij} 
{\cal D}_k X^k$ and ${\cal D}_i \hat{A}^{ij}_{\mathrm{TT}} = 0$, can be 
considered motivated by the local uniqueness properties of the elliptic sector 
shown in Ref.~\cite{CC09}.

We decompose classically the energy-momentum tensor, $T^{\mu\nu}$, measured by 
the observer of 4-velocity $n^\mu$ (Eulerian observer), in terms of the energy 
density $E := T_{\mu\nu} n^\mu n^\nu$, the momentum density 
$S_i := - \gm^\mu_i T_{\mu\nu} n^\nu$, and the stress tensor 
$S_{ij} := T_{\mu\nu} \gm^\mu_i \gm^\nu_j$, with $S := \gm^{ij} S_{ij}$ being 
its trace.

The resulting elliptic equations in the FCF are
\bea
	\tilde{\Delta} \psi &=& - 2 \pi \psi^{-1} E^* 
- \frac{\tgm_{il} \tgm_{jm} \hat{A}^{lm} \hat{A}^{ij}}{8 \psi^7} 
+ \frac{\psi \tilde{R}}{8}, \hspace{0.7cm} \label{e:psi} \\
	\tilde{\Delta} (N \psi) &=& \left[ 
2 \pi \psi^{-2} (E^* + 2 S^*) \right. \nonumber \\
	&&\left. + \left( 
\frac{7 \tgm_{il} \tgm_{jm} \hat{A}^{lm} \hat{A}^{ij}}{8 \psi^8} 
+ \frac{\tilde{R}}{8} \right) \right] (N \psi), \label{e:Npsi} \\
	\tilde{\Delta} \beta^i &=& 16 \pi N \psi^{-6} \tgm^{ij} (S^*)_j 
+ \hat{A}^{ij} {\cal D}_j (2N \psi^{-6}) \nonumber \\
	&&- 2 N \psi^{-6} \Delta^i_{kl} \hat{A}^{kl} \label{e:beta},
\eea
where the operator $\tilde{\Delta}$ is defined as 
\be
	\tilde{\Delta} \psi = \tgm^{kl} {\cal D}_k {\cal D}_l \psi
\ee
(analogously for $N \psi$) and
\be
	\tilde{\Delta} \beta^i = \tgm^{kl} {\cal D}_k {\cal D}_l \beta^i 
+ \frac{1}{3} \tgm^{ik} {\cal D}_k {\cal D}_l \beta^l,
\ee
$E^* := \psi^6 E$, $S^* := \psi^6 S$, $(S^*)_i := \psi^6 S_i$, 
\be
	\tilde{R} = \frac{1}{4} \tgm^{kl} {\cal D}_k h^{mn} {\cal D}_l \tgm_{mn}
- \frac{1}{2} \tgm^{kl} {\cal D}_k h^{mn} {\cal D}_n \tgm_{ml}
\ee
is the scalar 3-curvature of the conformal metric $\tgm_{ij}$, and
\be
	\Delta^i_{jk} = \frac{1}{2} \tgm^{kl} ({\cal D}_i \tgm_{lj} 
+ {\cal D}_j \tgm_{il} - {\cal D}_l \tgm_{ij})
\ee
is the difference between Christoffel symbols of the conformal and flat 
metrics.

The resulting hyperbolic equations are evolution equations for $h^{ij}$ and 
$\hat{A}^{ij}$,
\bea
	\partial_t h^{ij} &=& 2 N \psi^{-6} \hat{A}^{ij} + \beta^k {\cal D}_k h^{ij}
\nonumber \\
	&&- \tgm^{ik} {\cal D}_k \beta^j -  \tgm^{kj} {\cal D}_k \beta^i 
+ \frac{2}{3} \tgm^{ij} {\cal D}_k \beta^k, \label{e:hij} \\
	\partial_t \hat{A}^{ij} &=& (S_{\hat{A}})^{ij}, \label{e:hatAij}
\eea
where the explicit expression for the source $(S_{\hat{A}})^{ij}$ can be found 
in Ref.~\cite{CC12}.

If a TT decomposition is performed for $\hat{A}^{ij}$, an extra elliptic 
equation for the vector $X^i$ is added and Eq.~(\ref{e:hatAij}) can be viewed 
as an evolution equation for $\hat{A}^{ij}_{\mathrm{TT}}$.


\subsection{Spherical symmetry}
\label{sect:sphsym}
It has been proven in Ref.~\cite{CC11} that a spherically symmetric spacetime can be
locally foliated by a maximal slicing and by using isotropic coordinates for the 
spatial metric onto the spatial hypersurfaces $\Sigma_t$. This statement refers
to neighborhoods, and does not involve global prescriptions or boundaries. The 
FCF in spherical symmetry and with topologically $\mathbb{R}^3$ spatial 
hypersurfaces $\Sigma_t$ reduces to the isotropic gauge, where 
$\tgm^{ij} = f^{ij}$. This is not true anymore for more general topologies, 
like the $\mathbb{R}^3-{\cal B}$ case, where ${\cal B}$ is a ball, when general
boundary conditions on the boundary of ${\cal B}$ are given. In a nonconvex 
topology such as $\mathbb{R}^3-{\cal B}$, the expression for $\tgm^{ij}$ 
evolving in time in the FCF in spherical symmetry is instead given by
\be 
\tgm^{ij} (t) = \left( 
  \begin{tabular}{ccc}
    $\left( 1 + \frac{\omega(t)}{r^3} \right)^{\frac{2}{3}}$ & 0 & 0 \\
    0 & $\left( 1 + \frac{\omega(t)}{r^3} \right)^{-\frac{1}{3}}$ & 0 \\
    0 & 0 & $\left( 1 + \frac{\omega(t)}{r^3} \right)^{-\frac{1}{3}}$
  \end{tabular} \right),
\label{e:tgm-excision}
\ee
where $\omega(t)$ is a real and twice-derivable function of time $t$. A proof 
of this statement is presented in Appendix~\ref{app:tgmij}. Note that on 
$\mathbb{R}^3$, $\omega = 0$ is required at the origin for metric regularity. 
If a ball containing the origin is excised and the value for $\omega$ is not 
zero at the excision boundary, the spatial metric is not necessarily expressed 
as a conformally flat one.

Since the value for $\omega$ can be chosen arbitrarily taking into account the 
previous general expression for $\tgm^{ij}$ (it represents just a gauge 
freedom, as the Dirac gauge is a differential one), the spatial metric can be 
expressed as a conformally flat one, i.e., $\tgm^{ij} = f^{ij}$, or 
equivalently, $h^{ij} = 0$. Note that in this case Eq.~(\ref{e:hij}) is a 
time-independent prescription for $\hat{A}^{ij}$,
\be
	\hat{A}^{ij} = \frac{\psi^6}{2N} \left( \tgm^{ik} {\cal D}_k \beta^j 
+ \tgm^{kj} {\cal D}_k \beta^i - \frac{2}{3} \tgm^{ij} {\cal D}_k \beta^k 
\right),
\ee
Equations~(\ref{e:psi}--\ref{e:beta}) can be solved to obtain $N$, $\psi$ and 
$\beta^i$, and Eq.~(\ref{e:hatAij}) is a redundant condition in the bulk. This 
redundant condition will be used as a compatibility condition for the 
prescription of boundary conditions in the constrained system resolution.

In a general spacetime with no spherical symmetry $h^{ij} = 0$ cannot be 
imposed, and $h^{ij}$ and $\hat{A}^{ij}$ have to be evolved in time. Boundary 
conditions on the excision boundary for hyperbolic equations may or may not be 
imposed, depending on the characteristic structure of the system. This is not 
the case for elliptic equations, for which incorrect boundary conditions invalidate
the solution in the whole domain. The general case (no spherically symmetric 
spacetimes) is beyond the scope of this work and will be analyzed in future 
studies.


\section{Excision method}
\label{sect:excision}
Due to the singular character of BH interior solutions, measures have to be 
taken in numerical simulations of BH spacetimes. Quite a number of codes 
relying on hyperbolic formulations of Einstein's equations are based on an 
adaptive slicing which is typically designed to avoid the BH singularity by 
coordinate stretching and using a proper shift vector. This is the case of the BSSN 
formulation in combination with the puncture method, which is very popular in binary BH 
simulations. An alternative approach is known as stuffed BHs, where one fills 
BH interiors with unimportant (but regular) junk data in a hyperbolic 
formulation, and then evolves the regularized 
spacetime~\cite{arbona-98, brown-07}.

In this work, we want to present how the excision technique can be used in the 
FCF in spherically symmetric spacetimes, where the presence of elliptic PDEs 
has to be taken into account. This technique consists in removing from each 
spatial hypersurface $\Sigma_t$ the open interior of a topological sphere 
${\cal S}^2$, and solving Einstein's equations in the remaining hypersurface. The
sphere is assigned both physical and geometrical characteristics, tailored so 
that it is located strictly inside the AH of the modeled BH region, and so 
that it encloses the gravitational singularity. Those properties are then 
encoded as boundary conditions of the available elliptic equations to be solved
for.

In BH initial data problems, a natural approach consists in placing the 
excision sphere at the outermost marginally outer trapped surface in the 
initial slice, namely the AH. Much work has been done in this field, including 
how to impose this prescription and transcribe it in terms of 3+1 spacetime 
metric quantities (see, e.g., Refs.~\cite{CooPfe04, jaramillo-04, DaiJarKri04, 
gourgoulhon-06, CauCooGri06, Jaramillo:2009cc}). In the evolution case, the set
of excision spheres at every $\Sigma_t$ can be prescribed to describe a 
hypertube of marginally outer trapped surfaces, leading to a trapping horizon 
as outlined in Ref.~\cite{Jaramillo}. Here we will rather follow a more generic 
approach in which the excision surface is not enforced at the BH AH, but rather
at the interior of the AH world tube.

\subsection{Excision surface geometry}
We first define the geometrical setting of the excision surface. Notice that 
these definitions do not depend on the matter content of the spacetime, being 
valid for vacuum and nonvacuum spacetimes. Let ${\cal S}_t$ be a topological 
2-sphere embedded in $\Sigma_t$, and its induced 2-metric is $q_{ab}$. Let $s^\mu$
be the unit outward-directed spacelike vector normal to ${\cal S}_t$, that is 
also tangent to $\Sigma_t$. Let $\mathcal{H}$ be the hypertube formed by the 
set of excision spheres at every $\Sigma_t$. At the excision surface, three 
other vector fields are defined: the outward and inward future-directed null 
vectors
\be
	l^\mu = (n^\mu + s^\mu)/\sqrt{2}, \;\; k^\mu = (n^\mu - s^\mu)/\sqrt{2},
\ee
respectively, and the evolution vector on $\cal H$ normal to sections 
${\cal S}_t$ and carrying ${\cal S}_t$ onto ${\cal S}_{t+\delta t}$
\be
	h^\mu = N n^\mu + b s^\mu,
        \label{e:def_time_evol}
\ee
which we adapt to the 3+1 evolution vector $t^\mu = N n^\mu + \beta^\mu$, so 
that $b = \beta^i s_i$ (note that $b$ is only defined at $\cal H$). This 
means that the excision surface is kept at the same spatial coordinate 
location along the evolution. Two additional geometric quantities are the 
scalar outward expansion $\theta^{(l)}$ and outward shear $\sigma_{ab}$ along 
$l^\mu$, defined as
\bea
	{\cal L}_l \epsilon^{\cal S}_{ab} &=& \theta^{(l)}
        \epsilon^{\cal S}_{ab}, \label{e:def_theta_l}\\ 
	\sigma^{(l)}_{ab} &=& \frac{1}{2} \left[ {\cal L}_l q_{ab} 
- \theta^{(l)} q_{ab} \right],
\eea
where $\epsilon^{\cal S}_{ab}$ is the area element on ${\cal S}_t$. 
Analogously, the inward expansion $\theta^{(k)}$ and the corresponding shear 
can be defined.

For a Schwarzschild BH, in the case of adapting the excision surface to the AH,
the excision world tube $\cal H$ is a null hypersurface, meaning that the time 
evolution vector $h^\mu$ at the excision surface is null. This provides the 
following relationship between metric quantities at the AH: $b = N$. The 
outward expansion $\theta^{(l)}$ can be expressed as follows [see e.g. 
Eq.~(11.8) in Ref.~\cite{gourgoulhon-06}]:
\be
	\psi^2 \, \theta^{(l)} = 4 \tilde{s}^i \tilde{D}_i \ln \psi 
+ \tilde{D}_i \tilde{s}^i + K_{ij} \frac{\tilde{s}^i \tilde{s}^j}{\psi^2},
\label{e:expan_psi}
\ee
where $\tilde{s}^i := \psi^2 s^i$ and $\tilde{D}$ is the Levi-Civita connection
associated with the conformal metric $\tilde{\gamma}_{ij}$. This relation could
be used as a (nonlinear) Robin boundary condition on the conformal factor in 
order to compute initial data. In particular, the outward expansion 
$\theta^{(l)}$ vanishes if the excision surface is placed at the AH, or can be 
prescribed to be negative to place the excision surface inside the AH, as we 
shall do here. One can set $h^{ij} = 0$ at the boundary and on the whole 
spacetime, due to the particular form that the Dirac gauge takes in spherical 
symmetry (see Appendix~\ref{app:tgmij}).

The value of the lapse subsists at the excision boundary as a free condition 
for initial data: since the maximal slicing gauge provides us only with an 
elliptic constraint on the bulk, one can still choose freely the value at the 
inner excision surface. In the case of a dynamical evolution of a spherically 
symmetric BH, the value of the lapse at the excision surface has to be 
consistent with the assumption $h^{ij} = 0$, as it is described in the next 
subsection.

\subsection{Dynamical approach in the spherically symmetric case}
\label{sect:dyn_excision}
We consider now the time-dependent case in spherical symmetry involving matter 
content. Such situations include pure gauge evolutions (as illustrated in 
Sec.~\ref{ss:BH_evol}) and matter evolution (as illustrated here for the particular case of 
a scalar field, see Sec.~\ref{ss:accret}, and for the collapse of a neutron star to
a BH, see Sec.~\ref{ss:ns_bh}). As an astrophysical application, we have in 
mind the BH formation in stellar gravitational collapse simulations. Starting 
from a simulation for regular data evolving in $\mathbb{R}^3 \times [0, t_0]$, 
a trapped region forms at a given time $t=t_0$. To pursue the simulation long 
enough to study the subsequent evolution of the BH, one performs excision 
inside the trapped region, switching to a simulation in 
$(\mathbb{R}^3 - {\cal B}) \times [t_0, +\infty)$. The algorithm has the three 
following conditions to fulfill: i) for numerical stability reasons, a transition
between the two topologies has to occur smoothly, meaning that all the metric 
quantities solved for must be continuous and derivable in time at $t=t_0$; ii) 
dynamical excision has to avoid coordinate stretching and high gradient fields 
that would cause high and increasing inaccuracies in the computation; iii) the 
Schwarzschild solution should be recovered in the stationary limit.

The outermost trapped surface corresponding to the adopted spherically 
symmetric slicing can be located with an AH finder. Since there is no previous 
control on the geometry of this trapped region, the outermost trapped surface 
might be stretched or deformed in three-dimensional models, and thus is not in 
general an optimal candidate for the excision surface, unless we make an 
adaptation of the coordinates to the horizon that would imply a remapping of 
all data on the slice and likely introduce complexity in the problem and a 
copious amount of noise. A recent approach following this idea can be found 
in Ref.~\cite{Hemberger} for the simulation of binary BH spacetimes.

At time $t=t_0$, we choose the excision surface ${\cal S}_{t_0}$ to be located 
strictly inside the trapped region. The quantities $N$, $\psi$ and $\beta^i$ 
are determined at ${\cal S}_{t_0}$ by the previous evolution and are employed 
as initial values for the subsequent evolution. The outgoing scalar expansion 
is generically (and on average) negative, $\theta^{(l)}_{t=t_0} \leq 0$. 

Once the initial excised surface has been chosen, one needs to determine a 
geometrical prescription for the evolution of the excision surface in time, 
i.e. to characterize the excision hypertube. If the initial surface were the 
AH, one could prescribe it to span an AH world tube in time, by imposing the 
vanishing of the outward expansion $\theta^{(l)}$ at all times on the sphere of
constant radius $R$ \cite{Jaramillo}. Contrary to the stationary case, we do 
not have in general $b = N$ at the horizon. In particular, in the spherically 
symmetric case one has [see e.g. Eq. (38) in Ref.~\cite{Jaram11}, with 
$2C = b^2 - N^2$ and vanishing angular derivatives]
\bea
\label{e:b-N}
	\frac{b^2 - N^2}{2} = -
\frac{{\sigma^{(l)}_{\mu\nu}}{\sigma^{(l){\mu\nu}}} 
+ T_{\mu\nu}l^\mu l^\nu}{{\cal L}_k \theta^{(l)}} \ .
\eea
Under the null energy condition the numerator is nonpositive, so that the 
fulfillment of an outer trapping horizon condition \cite{Hayward:1993wb}, 
namely ${\cal L}_k \theta^{(l)}<0$, implies $b\geq N$. In this case the horizon
is either null (stationary case), or spacelike (dynamical case), depending on 
whether the energy flux vanishes across the BH horizon. Unfortunately, 
spherically symmetric trapping horizons do not necessarily fulfill the 
(stability) outer condition \cite{Booth:2005ng}, so that \textit{a priori} we 
cannot guarantee in general that $\cal H$ is spacelike in the dynamical case. 
This, together with the desire to avoid a coordinate adaption of the excision 
surface to the AH, leads us to choose an excision sphere strictly inside the AH 
and look for an appropriate characterization of the excision world tube. For 
instance, one could also impose a (nonpositive) value of the expansion 
throughout the evolution, which would also determine a hypertube geometry.

Here we will rather follow an effective approach in which we control the radial
component $b$ of the evolution vector $h^\mu$ on an excised coordinate sphere 
located strictly inside the AH. From 
$b = \beta^i s_i = \psi^2 \beta^i \tilde{s}_i$, it follows in spherical 
symmetry that $b = \beta^r \psi^2$. The imposition of a constant value in time of 
$b$ at the excised surface, given by the data at $t=t_0$, provides us with a 
simple boundary condition for the shift vector through time. We want the 
excision hypertube to be spacelike, so that the quantity $(b - N)$ should 
remain positive. Although we do not impose this condition directly, we monitor 
$b-N$ along the evolution so that $b$ could be dynamically adapted if needed to
guarantee the spacelike character of $\cal H$.

The values for $\psi$, $N$ and $h^{ij}$ have yet to be determined. The trace 
part of the evolution equations gives a consistent time evolution for $\psi$, 
valid everywhere and at all time [see, e.g., Eq.~(42) of Ref.~\cite{FCF}],
\be
	\partial_t \psi = \beta^k {\cal D}_k \psi + \frac{\psi}{6} {\cal D}_k \beta^k.
\label{e:evol-psi}
\ee
This equation, following from the kinematic definition of the extrinsic 
curvature, provides an additional coherent boundary condition for the conformal
factor, which is obtained by solving the corresponding (elliptic) 
Eq.~(\ref{e:psi}) with this boundary condition at the excised surface. The 
value of the lapse at the excised surface is the last (gauge) freedom left in 
the algorithm. We address this issue by making use of the form for $\tgm^{ij}$ in 
a $(\mathbb{R}^3 - {\cal B})$ topology in Eq.~(\ref{e:tgm-excision}). In 
particular, adopting the gauge  $\omega(t)=0$ fixes the remaining degree of 
freedom in the system and, in particular, fixes the value of the lapse on the 
excised surface. Indeed $\tgm^{ij}$ adopts then a conformally flat form (or 
equivalently $h^{ij}=0$) at all times, and by using Eq.~(\ref{e:hatAij}) to update
the extrinsic curvature at the excised surface we can fix the boundary 
condition for the lapse from any nondegenerate component of Eq.~(\ref{e:hij}); for 
instance,
\be
	N = \frac{\psi^6 (L \beta)^{ij} s_i s_j}{2\hat{A}^{ij} s_i s_j}.
\label{e:evol-lapse}
\ee

Our strategy can be summarized as follows: updated values for $\psi$ and 
$\hat{A}^{ij}$ at the excised surface are obtained by solving 
Eqs.~(\ref{e:evol-psi}) and (\ref{e:hatAij}), respectively; the imposition of 
constant $b$ and Eq.~(\ref{e:evol-lapse}) are used to obtain updated values of 
$N$ and $\beta^i$ at the excised surface; finally, $h^{ij}$ is vanishing 
throughout the evolution. During the numerical simulation, we check that 
$(b - N)\geq 0$ (see Sec.~\ref{sect:num}); under the choice of an excision 
world tube closely tracking the AH from its interior, this quantity is indeed 
expected to be non-negative for matter satisfying standard energy conditions in
stellar collapses [note however that in scenarios  not considered here that 
involve much larger BHs, the denominator in Eq.~(\ref{e:b-N}) can actually 
change sign, cf. \cite{Booth:2005ng}]. With these boundary conditions, all 
elliptic equations can be solved in the numerical domain for all times, and no 
evolution equations are solved in the bulk.

In this approach for using the excision technique we have not considered the TT
decomposition of the conformal extrinsic curvature given by 
Eq.~(\ref{e:hatAij}), motivated by a uniqueness pathology of the elliptic 
sector. We find in our numerical simulations of spherically symmetric 
spacetimes that the given boundary conditions at the excised surface for 
solving the elliptic equations are enough to avoid any convergence problem in 
the numerical resolution of the elliptic sector. However, this question is open
for more general spacetimes. In any case, the value of the $X^i$ vector of the 
TT decomposition of the conformal extrinsic curvature at the excised surface is
actually a degree of freedom~\cite{IsaCoconutM2009}.

\subsection{Convergence to a stationary solution}
\label{ss:analysis}
Let us comment here about the convergence of metric fields to stationary values
in our approach. Indeed, we find in our numerical simulations that the metric 
exponentially converges to a stationary solution. This fact means that the 
foliation induced by the boundary conditions described in 
Sec.~\ref{sect:dyn_excision} is such that the coordinates adapt to the 
stationarity of the spacetime. Since no evolution equations for $h^{ij}$ and 
$\hat{A}^{ij}$ are computed in the bulk in this approach, and solutions to 
elliptic equations are determined by the boundary conditions at the excised 
surface, we should focus on the analysis of the values of the metric variables 
at the excised surface. The value for $b$ is fixed to be constant at the 
excised surface, so a convergence of the conformal factor at the excised 
surface to a stationary value will imply also a convergence of the shift vector
at the excised surface to a stationary value. The evolution of the lapse $N$ at
the excised surface should be such that it is compatible with the setting 
$h^{ij}=0$ in the whole spacetime. This setting should in turn be compatible 
with coordinates adapted to stationarity.

We therefore focus on the value of the conformal factor $\psi$ at the excised 
surface, whose evolution is governed by Eq.~(\ref{e:evol-psi}). Let us define 
$\hat{\beta} = \beta^r \psi^2$, which coincides with $b$ at the excised 
surface. Equation~(\ref{e:evol-psi}) can be rewritten in terms of $\psi$ and 
$\hat{\beta}$ as
\be
	\partial_t \psi = \frac{\partial_r \hat{\beta}}{6 \psi} 
+ \frac{2\hat{\beta}}{3} \frac{\partial_r \psi}{\psi^2} 
+ \frac{\hat{\beta}}{3r \psi}.
\label{e:evol-psi2}
\ee

Here we keep $b$ constant at its initial given value at the excised 
surface, located at a fixed coordinate radius, say $R>0$, during the evolution:
$b_{|R} = \hat{\beta}_{|R} = b_0 > 0$ [since in the initial data $N$ is 
strictly positive and $(b-N)$ is positive, $b_0$ is strictly positive, too].
 
Motivated by the results we observe in our numerical simulations, let us assume 
in the present consistency analysis that $(\partial_r \hat{\beta})_{|R}$ is 
negative and does not change significantly during the evolution, so 
$(\partial_r \hat{\beta})_{|R} \approx b_1 < 0$, with $b_1$ being a constant. These 
assumptions are compatible with the ones found in our numerical simulations 
with an excised surface strictly inside the AH. In particular, such hypotheses 
(constant and negative value for $b_1$) are checked in the different numerical 
simulations of Sec.~\ref{sect:num}.

Let us assume a profile for $\psi$ of the form
\be
	\psi \approx 1 + \frac{c(t)}{r^p}, \;\; p \geq 1.\label{e:hyp_psi}
\ee
This profile can be considered as the one containing the leading term for $r$, 
taking into account that $\psi \to 1$ when $r \to \infty$ (first term) and that
the conformal factor should diverge at the center of the BH ($p \geq 1$). 
Therefore, $\partial_r \psi \approx -p \; c(t) / r^{p+1}$ and
$\partial_t \psi \approx c'(t) / r^p$, where the prime denotes the derivative. 

The hypotheses assumed here are not imposed during the numerical evolution of 
the system; they are used only to analyze the behavior of the metric variables 
in stationary spacetimes.

Taking into account previous assumptions, Eq.~(\ref{e:evol-psi2}) at the 
excised radius $R$ is rewritten as
\be
	c'(t) = \frac{c_1}{[R^p+c(t)]} + \frac{c_2}{[R^p+c(t)]^2},
\label{e:evol-c}
\ee
where
\bea
	c_1 &=& \frac{R^{2p-1}}{3} \left( \frac{b_1 \, R}{2} - b_0 (2p-1) \right) 
< 0, \label{e:def_c1}\\
	c_2 &=& \frac{2 b_0 \, p \, R^{3p-1}}{3} > 0,\label{e:def_c2}
\eea
are constants. We can integrate the previous differential equation, expressed in an
implicit way,
\bea
	t = f(R^p+c(t)) &=& c_0 + \frac{[R^p+c(t)]^2}{2c_1} 
- \frac{c_2 [R^p+c(t)]}{c_1^2} \nonumber \\
	&&+ \frac{c_2^2}{c_1^3} \log |c_1 [R^p+c(t)] + c_2|,
\eea
where $c_0$ is an integration constant. The implicit equation for $[R^p+c(t)]$ 
is well defined in both branches $c_1 [R^p+c(t)] + c_2 < 0$ and 
$c_1 [R^p+c(t)] + c_2 > 0$. The function $f$ has a monotonic behavior on each 
branch (increasing or decreasing depending on the specific one). In both 
branches, the range of $t$ is $\mathbb{R}$. For the initial value 
$[R^p+c(t=t_0)] = -c_2/c_1$, the solution of Eq.~(\ref{e:evol-c}) is simply the
constant value $[R^p+c(t)] = -c_2/c_1$. For initial values that are close by, an 
exponential convergence of $[R^p+c(t)]$ to the value $-c_2/c_1$ is found when 
$t \to +\infty$.

This discussion shows, under the assumed conditions, that we shall obtain 
exponential convergence to stationary values for the conformal factor $\psi$, 
and therefore to all metric variables, if the initial data are close enough to 
the stationary solution. This fact is checked numerically in 
Sec.~\ref{sect:num}.


\section{Numerical Results}
\label{sect:num}
The excision technique has been used in several numerical codes with different 
coordinates and slicings (e.g., in Ref.~\cite{Marsa96} a null-based slicing using 
minimally modified ingoing Eddington-Finkelstein coordinates was used to 
evolve a BH using the excision technique almost 20 years ago). In order to 
illustrate that the excision method presented in Sec.~\ref{sect:excision} for 
the FCF, in which maximal slicing and Dirac generalized gauge are chosen, works
well in practice in a numerical code, we have implemented and tested it in two 
toy models, detailed in the following subsections: the setup is presented in 
Sec.~\ref{ss:num_setup}, and it is applied to the evolution of a Schwarzschild BH in 
Sec.~\ref{ss:BH_evol}, and to the accretion of a massless scalar field in 
Sec.~\ref{ss:accret}. We have also implemented the excision technique in a full
simulation of the collapse of a neutron star to a BH in spherical symmetry in 
Sec.~\ref{ss:ns_bh}.

\subsection{Setup}\label{ss:num_setup}
We consider spherically symmetric spacetimes in the three following
physical scenarios: the evolution of a slicing of a Schwarzschild BH
(vacuum spacetime) in Sec.~\ref{ss:BH_evol}, the spherical accretion
of a massless scalar field onto an existing BH in Sec.~\ref{ss:accret}
and the collapse of an unstable neutron star to a BH. In all
cases, we use spherical (polar) coordinates and in the first two
scenarios, we start from an existing BH, with the excision sphere
located at the coordinate radius $r=1$. As stated in
Sec.~\ref{sect:sphsym}, the hypothesis of a spherically symmetric
spacetime in which adapted boundary conditions are chosen, such that
$\omega(t) = 0$ implies that, as far as the gravitational field is
concerned, we only need to solve for $N$, $\beta^i$ and $\psi$.

In the first two scenarios we start from initial data which already contain a BH 
and then evolve the system in the above-described excision scheme. Therefore we
need to construct initial data at $t=t_0=0$ by solving the elliptic 
system~(\ref{e:psi})-(\ref{e:beta}), with the stress-energy tensor being either
zero (vacuum) in the Schwarzschild BH evolution, or given by 
expressions~(\ref{e:E_phi})-(\ref{e:Sij_phi}) of 
Appendix~\ref{app:massless-scalar-field} in the case of the scalar field 
accretion. The initial excision surface is chosen as a sphere with a given 
(arbitrary) value of the outward expansion~[Eq.~(\ref{e:def_theta_l})] 
$\theta^{(l)}(t=0, r=1) = \theta^{(l)}_0$, prescribed to be negative. This 
guarantees that the initial excision surface is inside the AH, which then is 
located using an AH finder. Initial data boundary conditions are needed for the
elliptic system and are taken as follows:
\begin{itemize}
\item Setting $\theta^{(l)}_0$ and using Eq.~(\ref{e:expan_psi}), a 
(nonlinear) Robin boundary condition is obtained for the conformal factor 
$\psi$.

\item The boundary value of the lapse $N$ is fixed yielding a Dirichlet 
condition.

\item A value is prescribed for the quantity $b-N$ [see 
Eq.~(\ref{e:def_time_evol})], from which one can get a Dirichlet boundary 
condition for the radial component of the shift $\beta^r$, the other two 
components ($\beta^\theta, \beta^\varphi$) being zero in spherical symmetry.
\end{itemize}
The elliptic system giving the metric functions is solved iteratively, starting
from a first guess (flat metric) and inverting linear Laplace operators with 
the library \textsc{lorene}~\cite{lorene}, using multidomain spectral methods,
with a coordinate transform $u=1/r$ in the last domain, extending to infinity 
and allowing for the imposition of boundary conditions at $r \to \infty$ (see, 
e.g. Ref.~\cite{grandclement-09}).

In the third scenario (neutron star collapse to a BH), initial
data are obtained solving Einstein's equations, coupled to the fluid
equilibrium equations in the isotropic gauge, in whole space. The
numerical approach is very similar to the isolated BH case, but with no
inner boundary condition imposed. 

During the evolution, metric quantities are obtained by the resolution
of the same elliptic system as for the BH initial data, but
with different boundary conditions. As described in
Sec.~\ref{sect:dyn_excision}, the following boundary conditions are
used during the dynamical evolution:
\begin{itemize}
\item A Dirichlet boundary condition for the conformal factor $\psi$ is 
obtained from the time integration of Eq.~(\ref{e:evol-psi}) at the excision 
surface. As this surface is a sphere, there is no need for a boundary condition
to integrate Eq.~(\ref{e:evol-psi}) in time.
\item The value of the lapse at the boundary is set by the Dirichlet 
condition~(\ref{e:evol-lapse}). To compute it, we need to integrate in time 
Eq.~(\ref{e:hatAij}) at the excision surface.
\item The radial component of the shift is imposed at the excision boundary by 
keeping the value of $b(t)$ constant in time. 
\end{itemize}
The time integration of Eqs.(\ref{e:evol-psi})-(\ref{e:evol-lapse}) is done with a 
second-order (explicit) Adams-Bashforth scheme. 

\begin{figure*}
 \centering
  \includegraphics[height=0.48\textwidth,angle=-90]{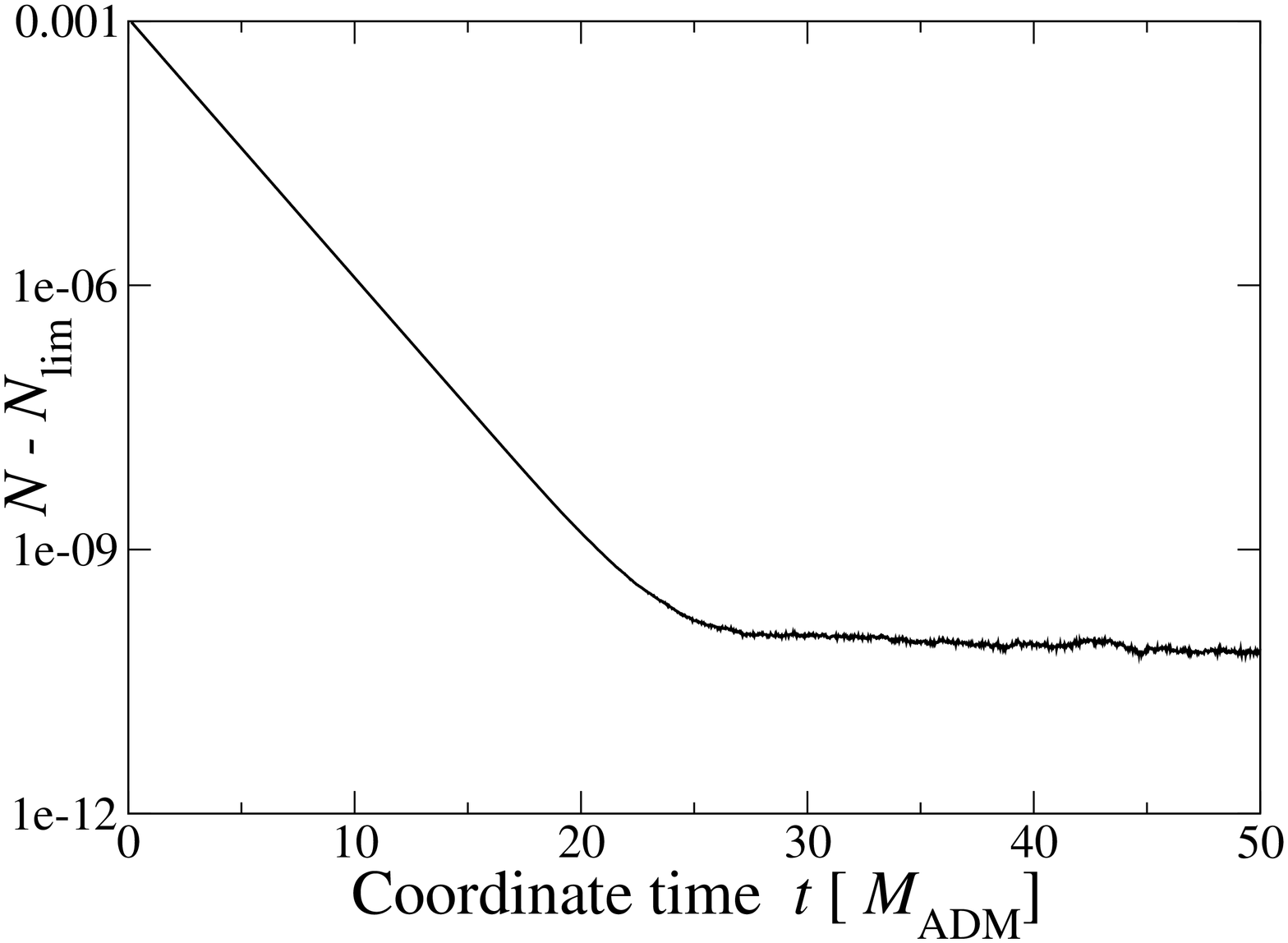}
  \includegraphics[height=0.48\textwidth,angle=-90]{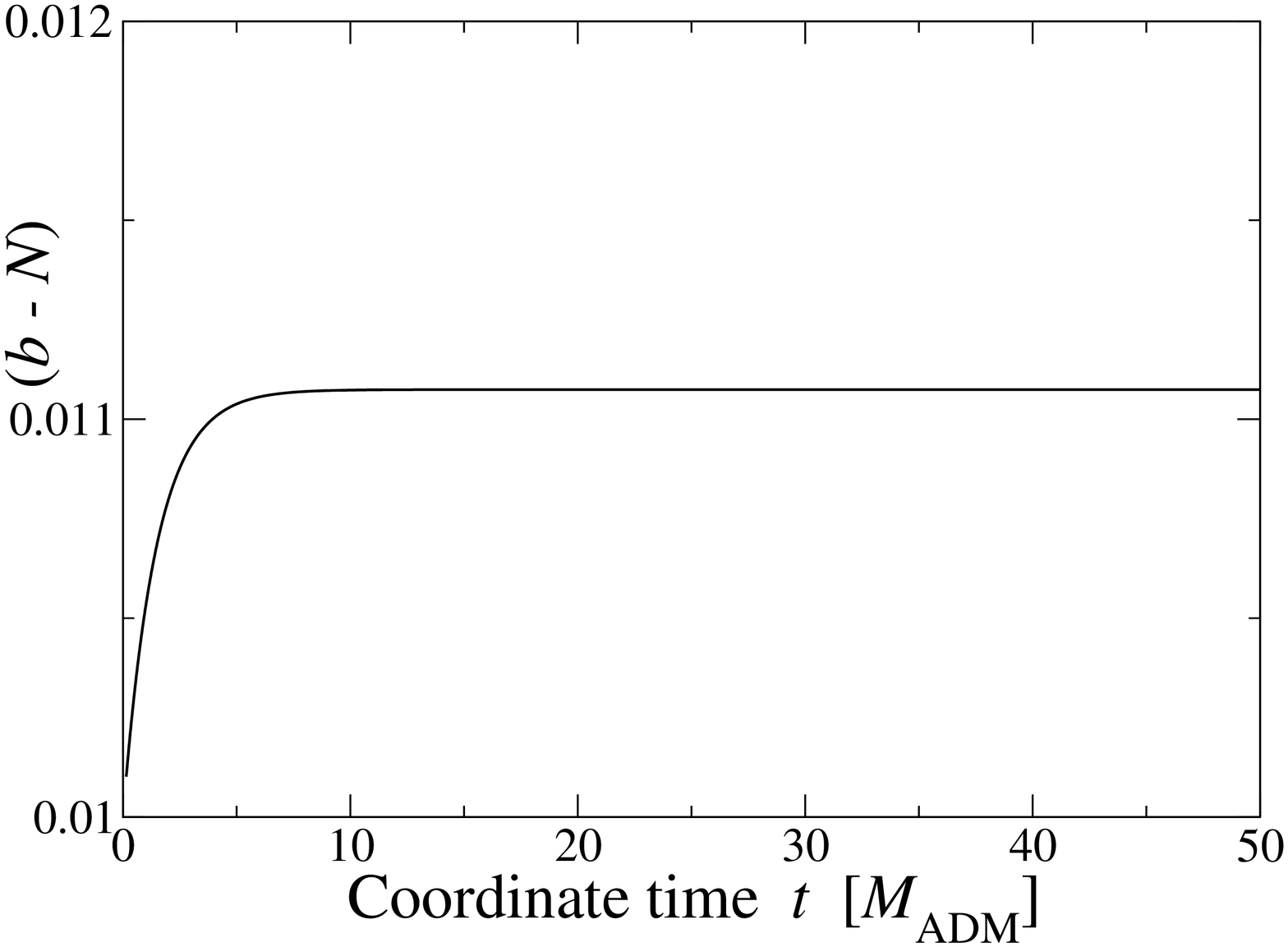}
  \includegraphics[height=0.48\textwidth,angle=-90]{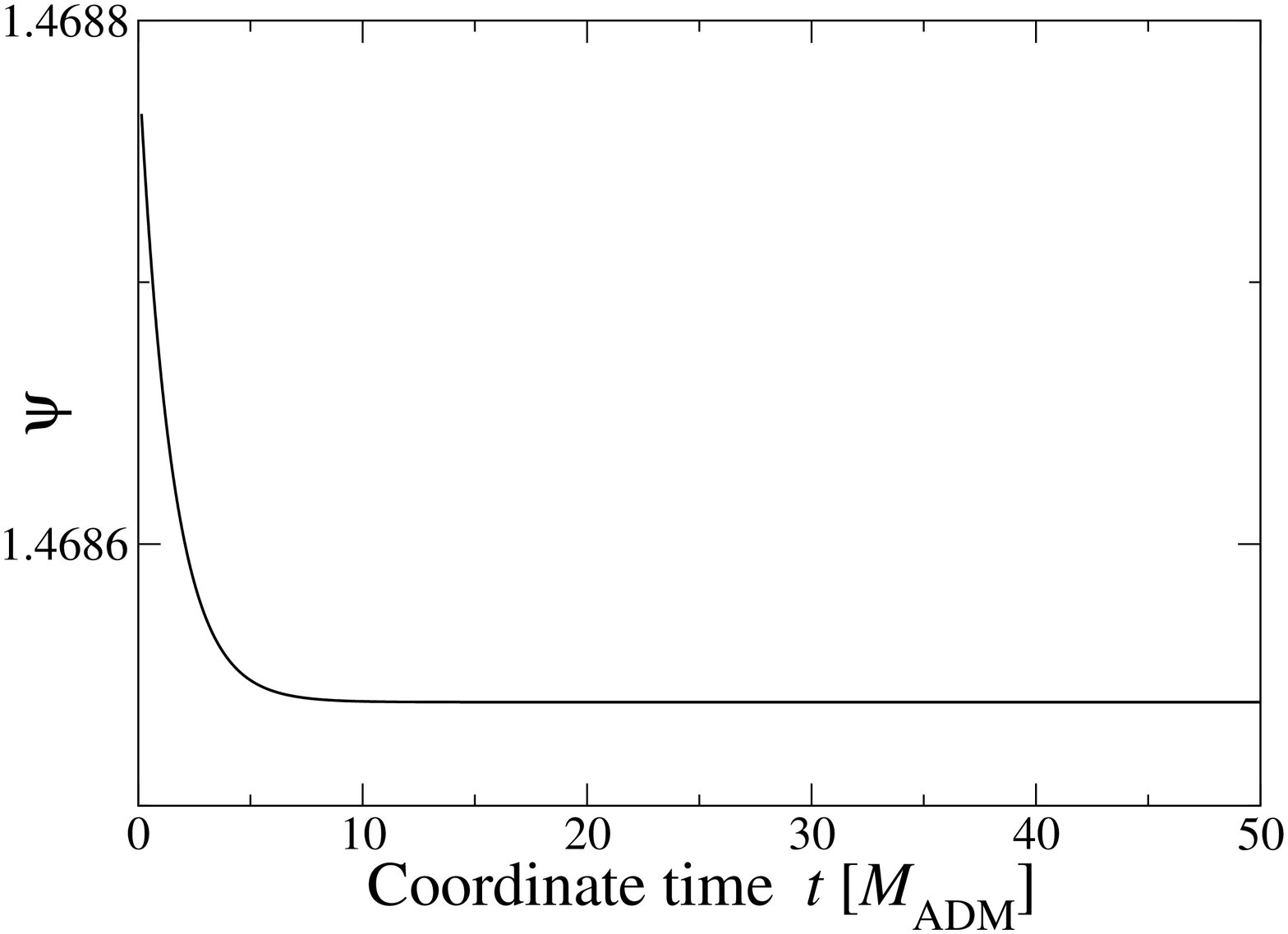}
  \includegraphics[height=0.48\textwidth,angle=-90]{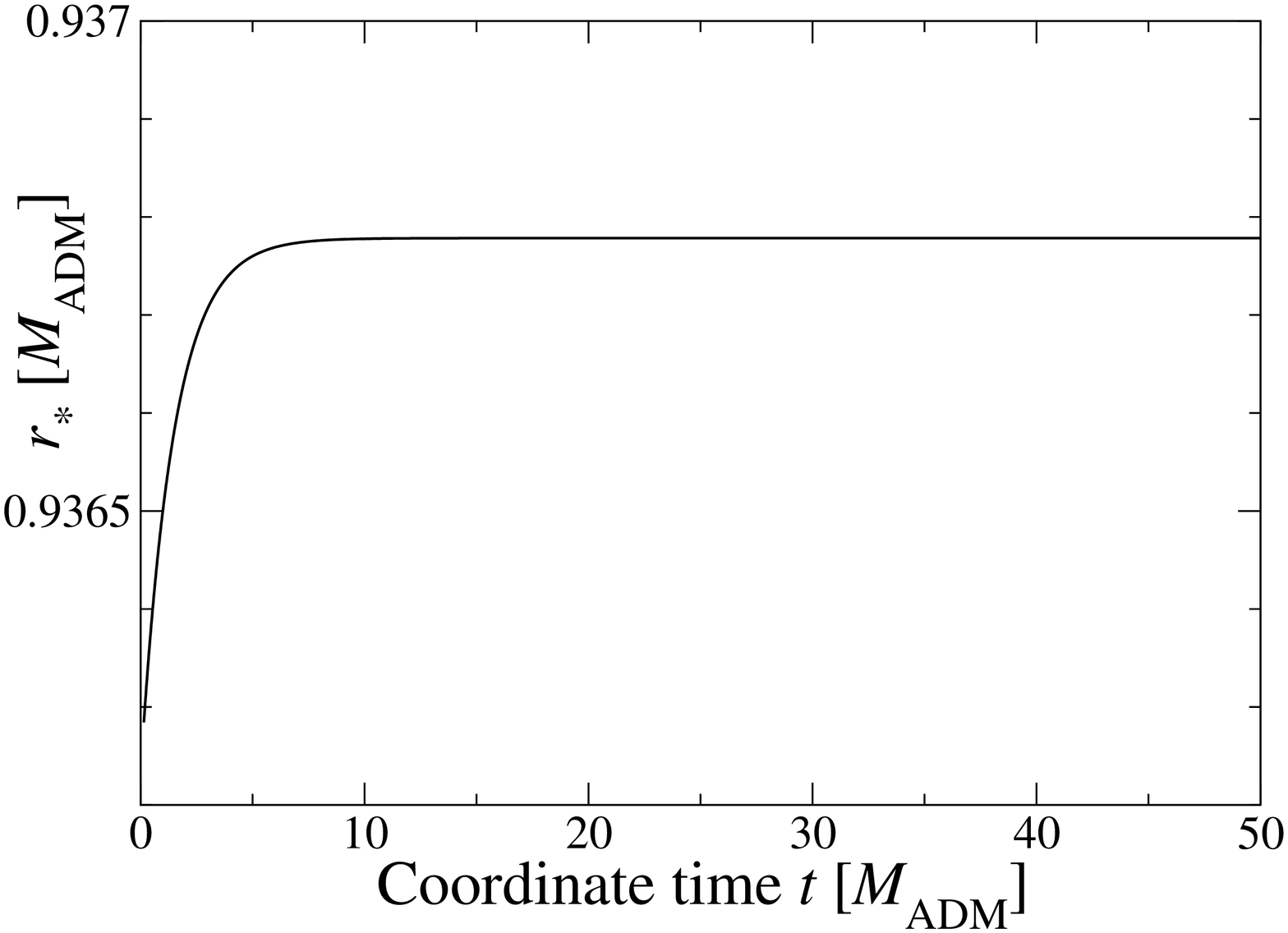}
  \caption{Evolution in terms of coordinate time $t$ of the lapse $N$, minus 
		its time asymptotic value $N_{\rm lim} \simeq 0.549$ (left top panel, 
		logarithmic scale); of $b$, the projection of the shift $\beta^i$ onto the 
		normal to the excision surface [Eq.~(\ref{e:def_time_evol})], minus the 
		lapse $N$ (right top panel); of the conformal factor $\psi$ (left bottom 
		panel), at the excision surface ($r=0.916\ M_{\rm ADM}$); and of the AH coordinate
		radius $r_*$ (right bottom panel), for a spherically symmetric vacuum (i.e.
		Schwarzschild) BH spacetime, using excision boundary conditions described 
		in Sec.~\ref{sect:dyn_excision}.}
\label{f:bh_metric_rAH}
\end{figure*}

\subsection{Evolution of a Schwarzschild black hole}\label{ss:BH_evol}
A spherically symmetric spacetime is computed for $r\geq 1$, with the previously 
defined boundary conditions, and the specific values $\theta^{(l)}_0 = -0.01$,
$N_{(t=0, r=1)} = 0.55$ and $\left(b - N\right)_{(t=0)} = 0.01$ for these 
initial data. Once having solved the elliptic system 
(\ref{e:psi})-(\ref{e:beta}), one verifies that this setting induces a nonzero
value for the Arnowitt-Deser-Misner (ADM) mass, more precisely 
$M_{\rm ADM} \simeq 1.09$. Numerically, it is obtained in isotropic gauge from 
the asymptotic behavior of the conformal factor (see e.g. Ref.~\cite{FCF}),
\begin{equation}
  \label{e:def_MADM}
  M_{\rm ADM} = - \frac{1}{2\pi} \oint_\infty \mathcal{D}_i \psi \, dS^i,
\end{equation}
where the integral is taken over a sphere of radius $r \to +\infty$. This causes
the excision sphere to be located at $r \simeq 0.916\ M_{\rm ADM}$. Using the 
numerical AH finder described in Ref.~\cite{lin-07}, we have found in the initial 
data an AH located at the coordinate radius $r_*\simeq 0.94\ M_{\rm ADM}$. 
This is evidence that the initial data represent a BH in spherical symmetry.

This BH spacetime is then numerically evolved for $t\geq 0$, through the time 
integration of boundary conditions, while solving the elliptic 
system~(\ref{e:psi})-(\ref{e:beta}) at every given number of time steps, as 
described in Secs.~\ref{sect:dyn_excision} and \ref{ss:num_setup}. This is 
obviously only a gauge evolution, since the spacetime is Schwarzschild by 
construction. The time evolution of the metric variables at the excision 
surface ($r=1=0.916\ M_{\rm ADM}$), namely $N$, $\psi$ and $b-N$, as well as 
the coordinate radius $r_*$ of the AH, in the interval $0 \leq t \leq 50$, are 
displayed in Fig.~\ref{f:bh_metric_rAH}. An exponential convergence toward 
stationary values is observed for all metric quantities, with an explicit 
behavior shown for the lapse $N$, as expected from the analysis carried out in 
Sec.~\ref{ss:analysis}. $r_*$ increases with time (right bottom panel of 
Fig.~\ref{f:bh_metric_rAH}), but the overall mass of the BH  does not (see 
below). 

\begin{figure}
 \centering
  \includegraphics[height=0.48\textwidth,angle=-90]{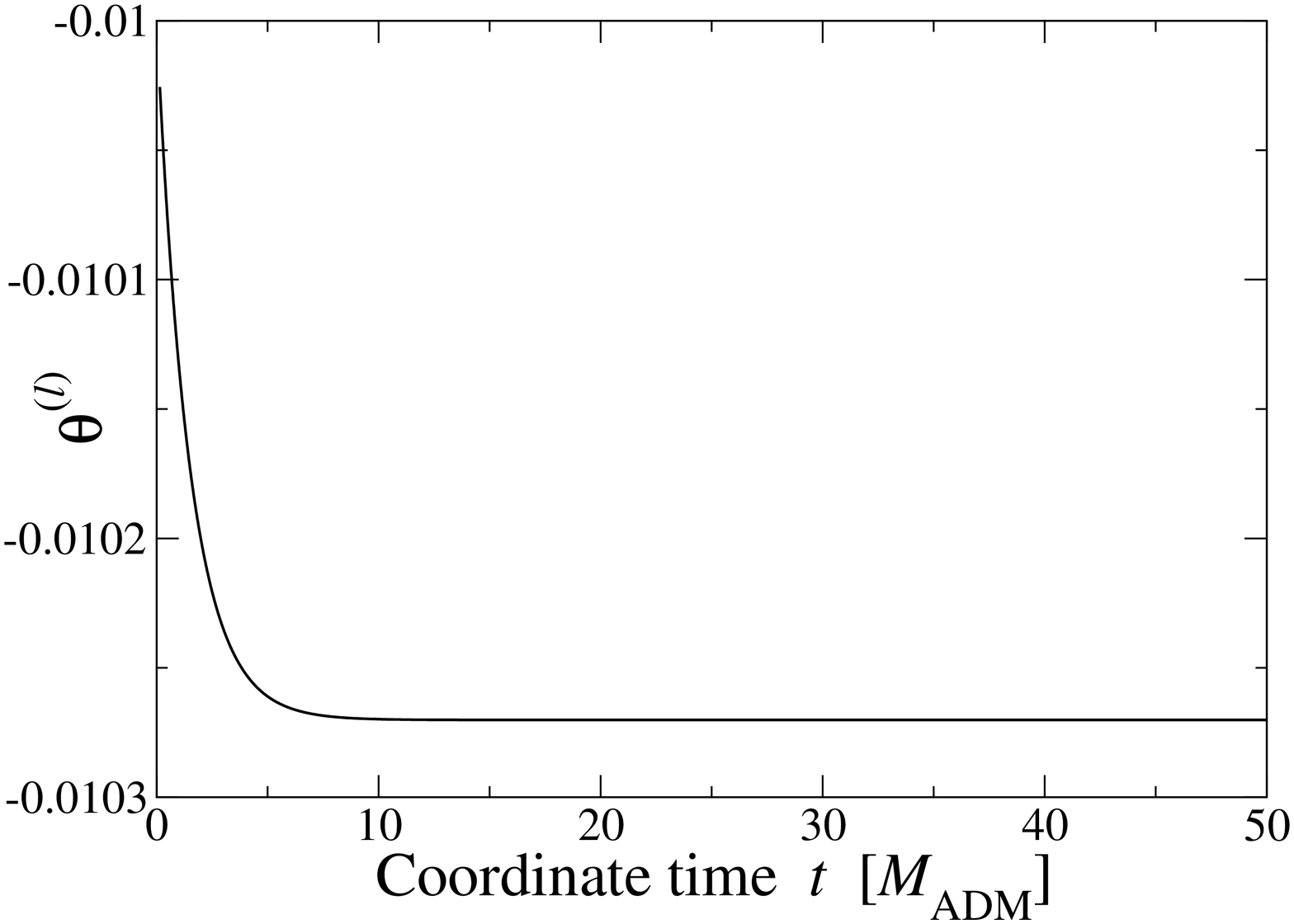}
  \caption{Evolution in terms of coordinate time $t$ of the outward expansion 
		$\theta^{(l)}$ [Eq.~(\ref{e:def_theta_l})], for a spherically symmetric BH 
		spacetime, as in Fig.~\ref{f:bh_metric_rAH}.}
\label{f:bh_theta}
\end{figure}

From the top right panel of Fig.~\ref{f:bh_metric_rAH}, one can check that the 
difference $b-N$ remains positive, as assumed in the discussion of 
Sec.~\ref{ss:analysis}. The hypothesis that $(\partial_r \hat{\beta})|_R$ is 
negative, assumed in the same analysis of Sec.~\ref{ss:analysis}, is also fulfilled 
during the evolution, and its value does not change significantly (the absolute
value of the relative difference with respect to its initial value is
$\lesssim 4 \cdot 10^{-3}$). Moreover, from the formulas (\ref{e:def_c1}) and 
(\ref{e:def_c2}), and the computed limit in Sec.~\ref{ss:analysis} for the 
conformal factor at the excision boundary, one can check that the 
hypothesis~(\ref{e:hyp_psi}) is valid; the numerically deduced value of $p$ 
from $\lim_{t\to\infty}\psi(t, r=1)$, and the values of $b_0$ and $b_1$, is 
$p\simeq 1.01$, which gives the expected behavior for the conformal factor. In 
Fig.~\ref{f:bh_theta} the time evolution of the outward expansion $\theta^{(l)}$
is displayed, which is decreasing during the simulation and always remains 
negative, ensuring that the excision surface always remains inside the AH.

\begin{figure}
 \centering
  \includegraphics[height=0.48\textwidth,angle=-90]{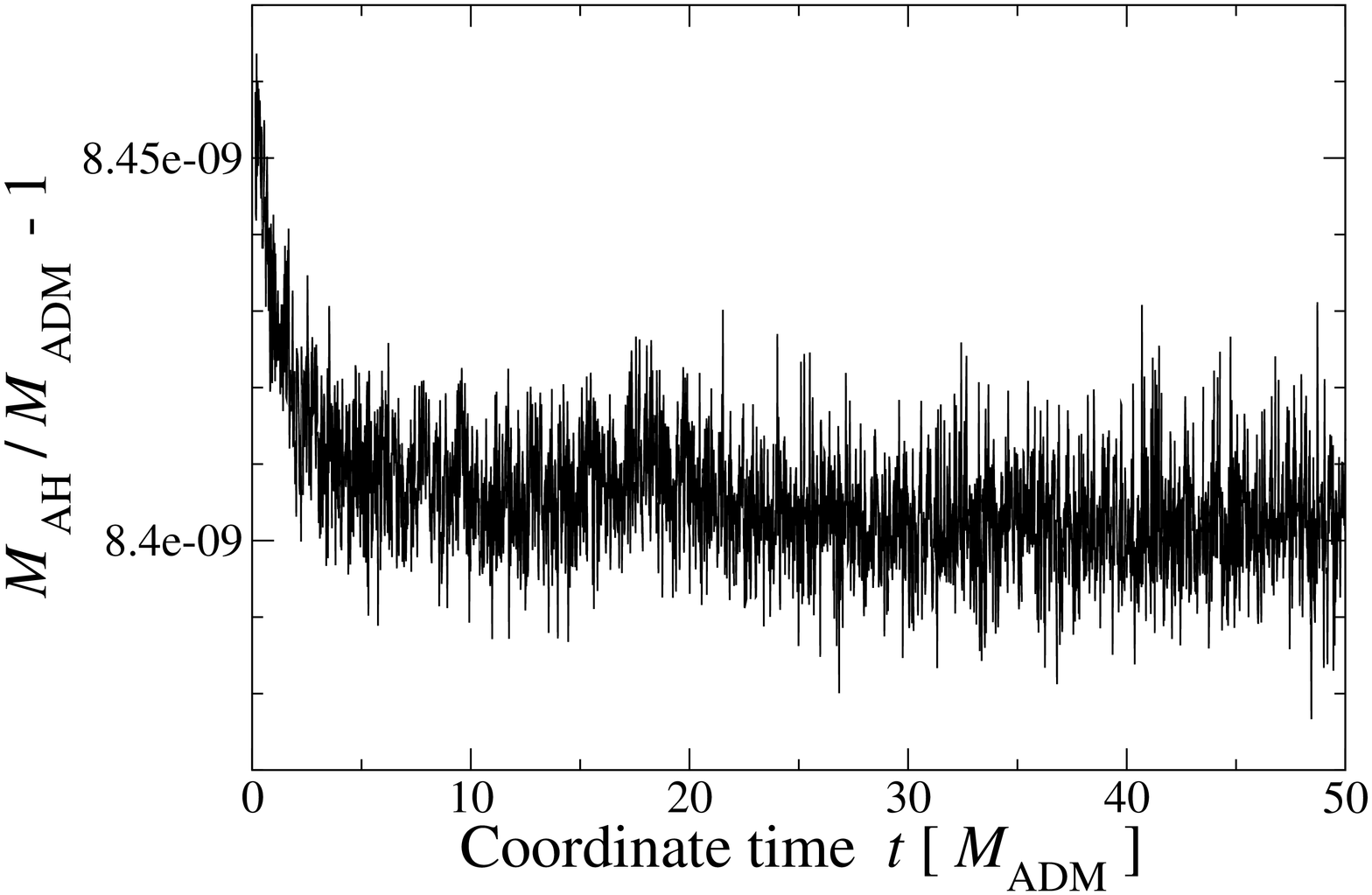}
  \caption{Evolution in terms of coordinate time $t$ of the variation of the AH
		irreducible mass~(\ref{e:def_rAH}) with respect to the initial 
		ADM mass~(\ref{e:def_MADM}), for a spherically symmetric BH spacetime, 
		using excision boundary conditions described in 
		Sec.~\ref{sect:dyn_excision}.}
\label{f:bh_rAH}
\end{figure}

The scheme appears stable (we have run it for $t \sim 1000$) and, in order 
to check its accuracy, we have monitored the variation of the AH irreducible 
mass, $M_{\rm AH}$, defined as
\begin{equation}
  \label{e:def_rAH}
  M_{\rm AH}(t) = \frac{1}{2}\sqrt{\frac{\mathcal{A}_{\rm AH}(t)}{4\pi}} =
  \frac{1}{2}\psi^2(t, r_*(t)) \, r_*(t),
\end{equation}
and determined by the AH finder. The Penrose inequality conjecture, in particular 
its rigidity part, provides a practical manner of characterizing the 
Schwarzschild solution. It states
\bea
	A\leq 16\pi M_{\rm ADM}^2 \Leftrightarrow M_{\rm AH} \leq M_{\rm ADM} \ ,
\label{e:Penrose}
\eea
where the equality is valid only for slices of the Schwarzschild spacetime. The 
conservation of the $M_{\rm AH}$ is displayed in Fig.~\ref{f:bh_rAH}, in 
particular showing the numerical consistency with a Schwarzschild solution 
using the Penrose inequality test. The conservation of the ADM 
mass~(\ref{e:def_MADM}) has also been checked, with quantitatively very similar
results to the conservation of the AH surface. Finally, a second-order 
convergence of these conserved quantities has been obtained numerically while 
decreasing the time step, in agreement with the implemented second-order 
Adams-Bashforth scheme.

The exponential convergence obtained for the chosen initial data, given by the 
specific values of $\theta^{(l)}_0$, $N_{(t=0, r=1)} > 0$ and 
$(b - N)_{(t=0)} > 0$, is independent of these initial data. A similar behavior
is found for different initial values.

\subsection{Accretion of a massless scalar field}\label{ss:accret}
In this case we evolve a BH spacetime with energy content in the form of a 
minimally coupled massless scalar field in spherical symmetry; see 
Appendix~\ref{app:massless-scalar-field} for details concerning the expressions
of the projections of the energy-momentum tensor, the corresponding evolution 
equations for the scalar field and the way they are solved. Initial data are 
given by a Gaussian profile for the scalar field outside the excision surface, 
i.e. for $r\geq 1$,
\begin{equation}
  \label{e:def_ini_phi}
  \phi(r,t=0) = \frac{\phi_0 r^2}{1+r^2}\,
  \left(e^{-(r-r_0)^2/\sigma^2}  + e^{-(r+r_0)^2/\sigma^2}\right),
\end{equation}
where $\phi_0$, $r_0$ and $\sigma$ are three constants, following 
e.g. Ref.~\cite{alcubierre-11}. The scalar field is evolved on the numerical grid up
to $r=R_{\rm max}= 120 \simeq 110\ M_{\rm ADM}$ (i.e. not in the compactified 
domain). Thus, the effect on the metric at the horizon of any scalar field wave
reflected from the artificial boundary at $r = R_{\rm max}$ can be in principle
neglected up to $t\sim 200$. For this simulation, we have used 24 numerical 
domains: one nucleus, 22 shells and a compactified domain (for details about 
the grid setting, see Ref.~\cite{grandclement-09}). As said, the wave equation is 
not solved in the compactified domain, but instead the outgoing boundary 
conditions~(\ref{e:BC_pi_phir})-(\ref{e:BC_phi}) are imposed at 
$r = R_{\rm max}$. Twenty-five Chebyshev coefficients are used in each domain, in order 
to describe the wave accurately enough.

\begin{figure}
 \centering
  \includegraphics[height=0.48\textwidth,angle=-90]{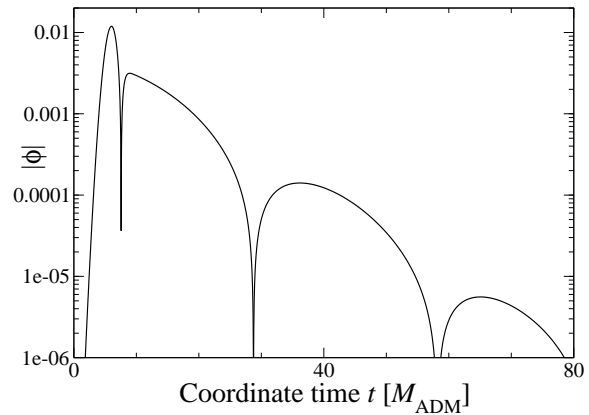}
	\caption{Evolution in terms of the coordinate time $t$ of the absolute value 
	  of the accreted scalar field, $|\phi|$, at the excision surface, 
		$\phi(r=1,t)$.}
\label{f:scal_phi}
\end{figure}

\begin{figure}
 \centering
  \includegraphics[height=0.48\textwidth,angle=-90]{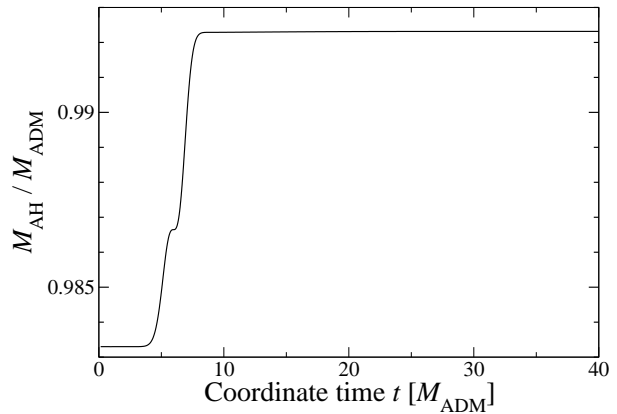}
  \caption{Evolution of the ratio between the AH irreducible
    mass~(\ref{e:def_rAH}) and the ADM one~(\ref{e:def_MADM}), as a
    function of the coordinate time $t$, for the case of the accretion
    of a massless scalar field.}
\label{f:scal_rAH}
\end{figure}

\begin{figure}
 \centering
  \includegraphics[height=0.48\textwidth,angle=-90]{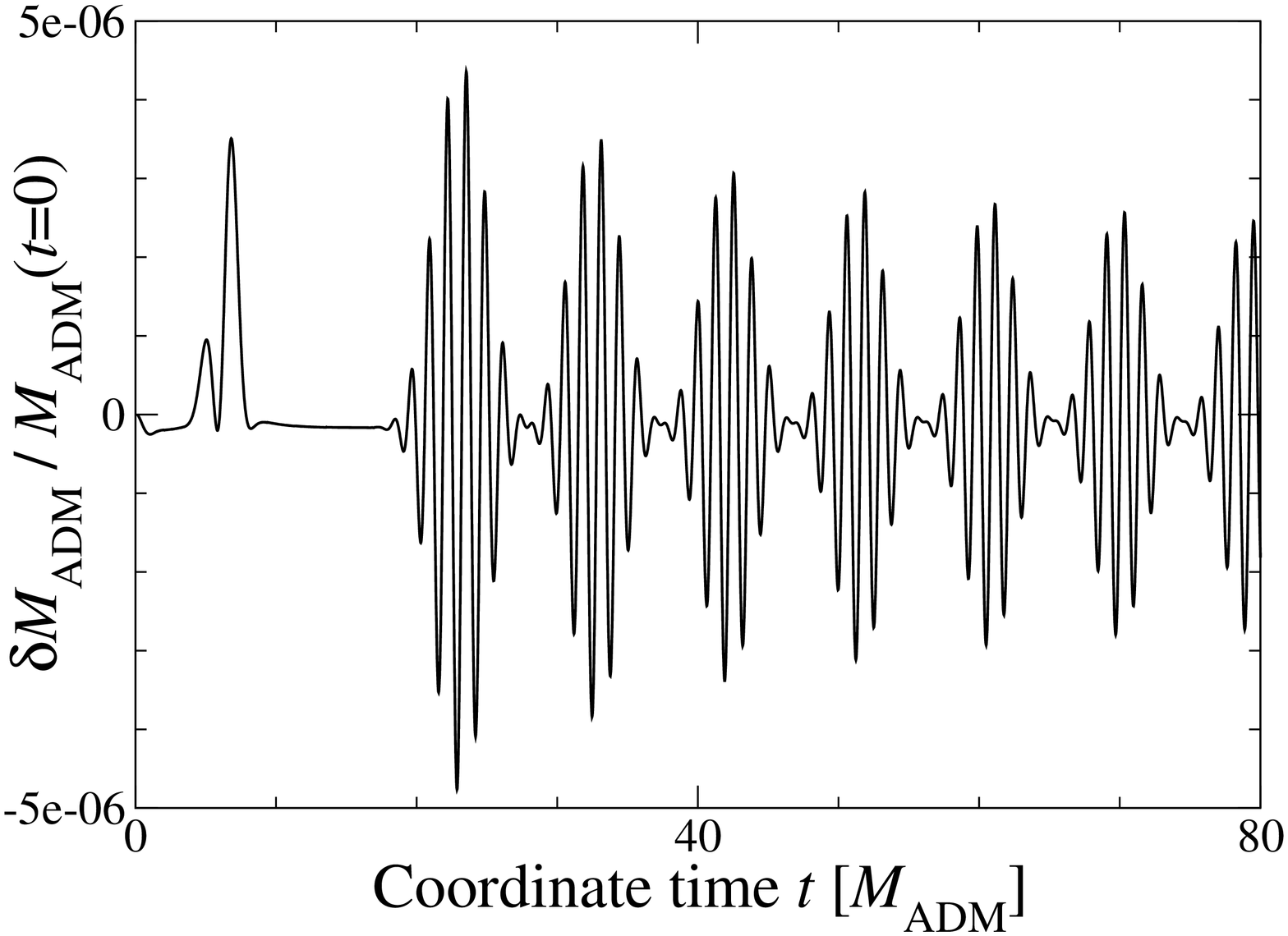}
  \caption{Evolution in terms of the coordinate time $t$ of the relative 
		variation of the ADM mass, defined as in Eq.~(\ref{e:def_MADM}), with 
		respect to its initial value, for the case of the accretion of a massless 
		scalar field.}
\label{f:scal_MADM}
\end{figure}

We evolve these initial data with $\phi_0 = 0.01$, $r_0 = 5$ and $\sigma = 1$. 
A fraction of the scalar field is radiated away, while the other part is accreted 
onto the BH; its time evolution at the excision surface is given in 
Fig.~\ref{f:scal_phi}. The metric quantities, $N$, $\beta^r$ and $\psi$ follow 
a similar evolution with respect to the vacuum case and, once the scalar field 
has been accreted to the BH, they settle rapidly to stationary values. 
Figure~\ref{f:scal_rAH} gives the evolution of the AH mass as a function of the 
proper time of the observer that is located at the AH. As expected, the AH 
grows in time while accreting energy from the scalar field, before reaching a 
stationary limit. This limit does not represent all of the ADM mass of the 
spacetime, as part of this asymptotic mass is still contained in the scalar 
field traveling to higher radii. We have checked the accuracy of the code by 
monitoring the variation of the ADM mass, computed by Eq.~(\ref{e:def_MADM}) in
Fig.~\ref{f:scal_MADM}. The conservation of this quantity, up to the level 
$10^{-6}$, shows that the scalar field stress-energy enters the BH and makes it
grow accordingly. The first two spikes for $t\lesssim 10$ can be attributed to 
the scalar field which is entering the BH AH. Further oscillations that are 
seen in this figure for $t\gtrsim 20$ can be related to the passing of the 
scalar field wave from one spectral domain to another. Note that the overall 
level of this ADM mass violation is $10^{-6}$ and that it converges away with 
both time and spatial resolutions. 

This simulation shows that the excision boundary conditions that have been 
designed here allow us to study the growth of a spherically symmetric accreting
BH in a stable and accurate way.

\subsection{Collapse of a neutron star to a black hole}\label{ss:ns_bh}

We here describe a simulation in spherical symmetry, starting from an unstable 
static neutron star, up to the formation of a BH and its accretion of all 
matter into the horizon. Excision is switched on during the simulation, after 
the AH is formed. For this simulation, we have modified the code 
\textsc{CoCoNuT}~\cite{dimmelmeier-05}, so that it uses the excision technique 
described above, in the case of spherical symmetry. This code solves the 
general-relativistic Euler equations (see Appendix~\ref{app:perfect_fluid} for 
details), with Einstein's equations in isotropic gauge. As far as hydrodynamics 
are concerned, in the case of grid boundary inside an AH, there is no need for 
boundary conditions as all the characteristics point out of the numerical 
integration domain (e.g. Ref.~\cite{hawke-05}). In practice, the Euler equations are
solved with the use of high-resolution shock-capturing schemes 
(see Ref.~\cite{font-08}) and, in order to compute the fluxes at the boundary, we 
perform a simple copy of primitive variables into the ghost cells: the rest-mass 
density $\rho$, 3-velocity $v^i$ and internal energy $\epsilon$ 
(see Ref.~\cite{dimmelmeier-05} for definitions). The metric equations are the same as 
in the previous numerical example, but with the matter sources 
($E, S_i, S_{ij}$) computed from the perfect-fluid stress-energy tensor 
obtained from the integration of Euler equations. Further details on the 
modeling of the neutron star collapse to a BH can be obtained in Ref.~\cite{CC09}.

\begin{figure}
 \centering
  \includegraphics[height=0.48\textwidth,angle=-90]{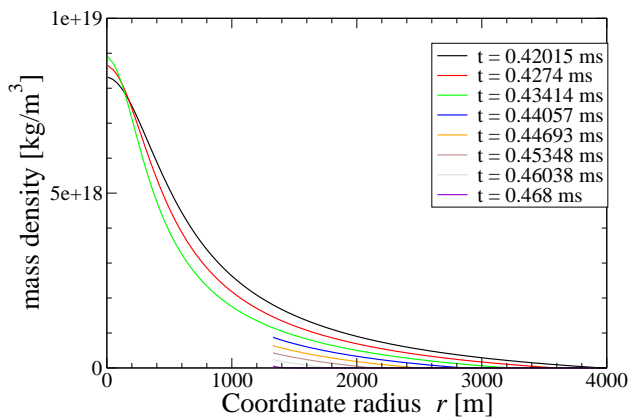}
  \caption{Radial profiles of the rest-mass density, for different moments 
		around the start of the excision in the simulation of the collapse of a 
		neutron star to a BH. Excision starts at $t=0.43788$~ms.}
\label{f:nsbh_rhovsr}
\end{figure}
 
The initial data for this simulation consists of a static neutron star, computed in 
isotropic gauge with a polytropic equation of state of adiabatic index 
$\gamma_{\rm eos} = 2$. The central density is such that the star lies on the 
unstable branch, i.e., it can either migrate toward a stable 
configuration with the same baryon number or collapse to a BH (see 
also Ref.~\cite{CC09}). The star has the following initial properties: 
gravitational (ADM) mass $M_{\rm ADM} = 1.617~M_\odot$, baryon mass 
$M_B = 1.771~M_\odot$, coordinate radius $R_{\rm star} = 7.825$~km and 
integrated (cicular) radius $R_{\rm circ} = 10.39$~km. On top of this 
equilibrium configuration, we have added a $1\%$ density perturbation, in order
to ensure that the star collapses to a BH, and does not migrate to the stable 
branch.

\begin{figure}
 \centering
  \includegraphics[height=0.48\textwidth,angle=-90]{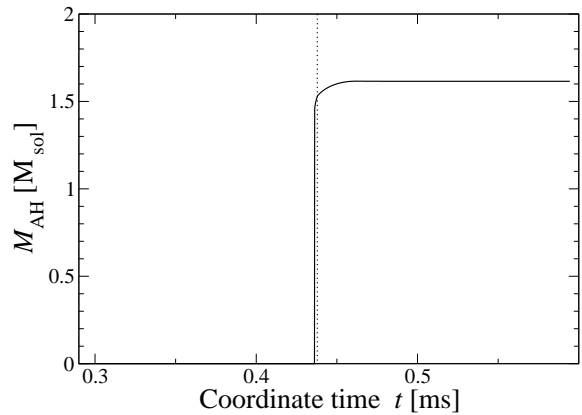}
  \caption{Evolution in terms of coordinate time $t$ of the AH irreducible 
		mass~(\ref{e:def_rAH}) for the collapse of a neutron star to a BH. The vertical
		dotted line indicates the moment when excision is turned on.}
\label{f:nsbh_MAH}
\end{figure}

The instant $t=0$ corresponds to the beginning of the collapse, which proceeds 
until the formation of an AH. Excision is switched on when a given ratio of the
total baryon mass has entered the horizon; in practice, we have set this ratio to 
$85\%$, but we have checked that changing this value had little influence on 
the evolution of observable quantities. Note that, if this ratio is small 
($\lesssim 80\%$), excision is switched on immediately after the detection of 
the AH. The excision radius is defined inside the AH radius $r_*$, with a value 
$r_{\rm excision} / r_*(t=t_{\rm excision}) \in [0.9,0.98]$. Again, it has been
checked that the choice for this ratio does not influence physical results. In 
the run shown here, $r_{\rm excision} = 1329$~m. Density radial 
profiles at various time steps around the time when excision is started 
($t=0.43788$~ms) are given in Fig.~\ref{f:nsbh_rhovsr}. In particular, the density 
distribution keeps a smooth behavior after the excision is switched on, and 
matter proceeds to fall into the BH, which reaches a stationary state, 
surrounded by vacuum. This is illustrated in Fig.~\ref{f:nsbh_MAH}, where we 
have plotted the time evolution of the BH irreducible mass, defined by 
Eq.~(\ref{e:def_rAH}). When the AH appears, it grows abruptly; the excision is 
then switched on and the BH accretes matter that was left outside the horizon, 
before settling down to the stable and stationary case. The whole simulation 
then remains stable, with no noticeable change in any evolved quantity.

\begin{figure}
 \centering
  \includegraphics[height=0.48\textwidth,angle=-90]{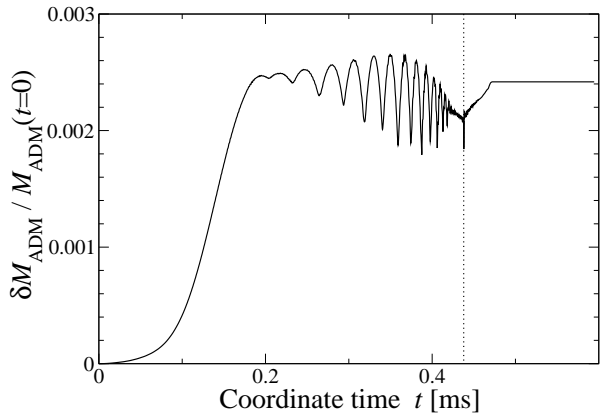}
  \caption{Evolution in terms of coordinate time $t$ of the relative variation,
		with respect to its initial value at $t=0$, of the ADM 
		mass~(\ref{e:def_MADM}) for the spacetime of a neutron star collapsing to a
		BH. The vertical dotted line indicates the moment when excision is turned on.}
\label{f:nsbh_ADM}
\end{figure}

We observe exponential convergence of metric quantities on the excision 
surface, as in Sec.~\ref{ss:BH_evol}, as soon as all matter has passed the 
excision surface. To test the accuracy of the technique, we plot the relative 
time variation of the ADM mass~(\ref{e:def_MADM}) of the spacetime in 
Fig.~\ref{f:nsbh_ADM}. The main error in the conservation of the ADM mass comes
from the numerical solution of the Euler equations, using finite-volume methods
(512 radial cells have been used). After excision has been started (dotted line
in Fig.~\ref{f:nsbh_ADM}), this error shows a small change, and remains almost 
constant when the accretion phase has ended. The subsequent relative change in 
the ADM mass is then of the order of $10^{-8}$, similarly to 
Sec.~\ref{ss:BH_evol}. Note that the conservation of the ADM mass is not enforced 
in the numerical scheme, but only monitored as a global test of the accuracy of
the code.
 

\section{Conclusions}
\label{sect:conclusion}
In this work we have presented a new excision technique for the dynamical 
evolution of spherically symmetric spacetimes in the FCF in order to 
numerically simulate systems forming a BH.

FCF belongs to the so-called constrained formulations of Einstein's equations, in
which the constraints are solved for each time step. On the contrary, in free 
evolution formulations the evolution equations are in general of hyperbolic 
type and constraints are used to monitor the validity of the numerical solution
(and/or as damping terms in the evolution scheme). The puncture method has been
used in combination with the BSSN formulation for binary BH evolutions and the 
excision technique in combination with the generalized harmonic gauge. The 
difficulty of using the excision technique in the case of constrained 
formulations comes from the fact that constraints are elliptic-type PDEs and 
incorrect boundary conditions at the excision surface invalidate the physical 
solution in the whole numerical domain.

In the context of constrained formulations, excision has been used to generate 
initial data. Dynamical evolutions using constrained formulations were 
presented in the work of Scheel \textit{et al.}~\cite{scheel-95}, in the case 
of dust collapse in Brans-Dicke theory of gravity; in their work, the excision 
boundary was considered to be the AH at a fixed radial coordinate. In 
nonspherically symmetric spacetimes, the coordinate shape of the AH can deviate 
from a coordinate sphere and this approach has some limitations. In our case, 
as in that of Rinne and Moncrief~\cite{rinne-13}, the excision boundary is a coordinate 
sphere located strictly inside the AH, and we let the coordinate location of 
the AH evolve freely in time. This approach allows in particular for a very 
straightforward extension to spacetimes without symmetries, where the AH can 
form with a shape deviating from a coordinate sphere.

The proposed approach uses an arbitrary coordinate sphere located strictly 
inside the AH as excision surface and a set of simple boundary conditions for 
the elliptic equations to be solved. It permits more freedom in the choice of 
the excision surface, which could be set as a simple coordinate sphere in 
spacetimes without symmetries. We have checked the practical applicability of 
this approach in three cases: the numerical simulation of a Schwarzschild 
spacetime, the spherically symmetric accreting BH with energy content in the 
form of a massless scalar field, and the collapse of a spherical neutron star 
to a BH. Our numerical results are stable and accurate. We have also 
theoretically analyzed the behavior of these boundary conditions in the 
proximity of stationary spacetimes and found an exponential adjustment of the 
coordinates to stationary values, independently of the chosen initial data. 
This behavior has also been checked numerically. In the last studied case, we 
have demonstrated that the switching on of excision during the collapse did not
introduce any additional noise and that the overall simulation remained stable,
with the newly formed BH accreting matter outside the excision surface. We can 
thus follow the whole BH formation process, from the onset of the collapse, to 
the growth of the BH and the description of the stationary solution up to 
arbitrarily long times. Although we have restricted this work to spherically 
symmetric spacetimes, we plan to extend this approach to more 
general spacetimes with less symmetries in forthcoming studies.

This excision technique can be used in the context of several astrophysical 
scenarios like a stellar collapse to a BH as we have shown, but also in other 
scenarios like the formation of an accretion disk and/or the launching of a 
jet. Notice that in these scenarios the initial data are usually regular, and 
during the simulation a BH forms and, particularly when using singularity-avoiding 
time coordinates (e.g. maximal slicing), an AH is found before the 
appearance of the physical singularity~\cite{shapiro-80}. Then the excision 
technique can be used to continue the numerical simulation, say in the 
accreting period, avoiding the stretching of the grid at the center.

An advantage of this technique in combination with the use of spherical (polar)
coordinates is the avoidance of the time-step limiting in the numerical 
simulations due to the small size of the central cells. This point can be more 
important for three-dimensional simulations.


\section*{Acknowledgments}
We thank P.~Montero and T.~Baumgarte for fruitful comments and discussions. 
I. C.-C. acknowledges support from the Alexander von Humboldt Foundation. This work
has been partially funded by the SN2NS project ANR-10-BLAN-0503. This work was 
also supported by the Grant No. AYA2010-21097-C03-01 of the Spanish MICINN.


\appendix

\section{Dirac gauge and spherical symmetry}
\label{app:tgmij}
Spherical symmetry allows us to restrict the general form of the conformal 
metric in Eq.~(\ref{e:Dirac}), expressed in orthonormal spherical 
coordinates, to
\be
	\tgm^{ij} = \left( \begin{tabular}{ccc} 
	                   $A(r)$ & 0 & 0 \\
	                   0 & $B(r)$ & 0 \\
	                   0 & 0 & $C(r)$
	                   \end{tabular} \right),
\ee
with the additional determinant condition
\be
	A(r) \, B(r) \, C(r) = \det f^{ij} = 1.
\ee
Equation~(\ref{e:Dirac}) and the previous one for the determinant can be written 
as
\bea
	&&\partial_r A + \frac{2A - B - C}{r} = 0, \;\; B = C = \frac{1}{\sqrt{A}} 
\nonumber \\
	&&\Leftrightarrow \partial_r A + \frac{2 (A - 1/\sqrt{A})}{r} = 0, \;\; 
B = C = \frac{1}{\sqrt{A}}.
\eea
A general solution for $A$ is
\be
	A(r) = \left( 1+ \frac{\omega}{r^3} \right)^{2/3}, \; \omega \in \mathbb{R}.
\ee

\section{Massless scalar field in a curved spacetime}
\label{app:massless-scalar-field}
The massless Klein-Gordon equation, or (simply) the wave scalar equation, is given by
\be
	\nabla^\mu \nabla_\mu \phi = 0,
\label{e:KG}
\ee
where $\nabla$ is the Levi-Civita connection associated with the spacetime 
metric $g_{\mu\nu}$. The stress-energy tensor associated with this scalar field
is given by
\be
	T_{\mu\nu} = \nabla_\mu \phi \nabla_\nu \phi 
- \frac{1}{2} \gamma_{\mu \nu} \nabla_\rho \phi \nabla^\rho \phi.
\ee
Its projections are given by
\bea
	E &=& \frac{1}{2N^2} \left( (\partial_t - {\cal L}_\beta) \phi \right)^2 
+ \frac{1}{2} D_\rho \phi D^\rho \phi, \label{e:E_phi}\\
	S_i &=& \frac{1}{N} \left( (\partial_t - {\cal L}_\beta) \phi \right) 
D_i \phi, \label{e:Si_phi}\\
	S_{ij} &=& D_i \phi D_j \phi \nonumber \\
	&& - \frac{1}{2} \gamma_{ij} \left[ D_k \phi D^k \phi 
- \frac{1}{N^2} \left( (\partial_t - {\cal L}_\beta) \phi \right)^2
\right], \label{e:Sij_phi}
\eea
where $D$ is the Levi-Civita connection associated with the spatial metric 
$\gamma_{ij}$.

The wave equation~(\ref{e:KG}) is rewritten as a first-order system in space
and time, by introducing the auxiliary scalar $\Pi$, defined from 
Eq.~(\ref{e:defPi}) below, and the vector $\Phi_i = {\cal D}_i \phi$, 
considered also as a constraint of the system, as
\bea
	\partial_t \phi &=& - N \, \Pi + \beta^i {\cal D}_i \phi, \label{e:defPi} \\
	\partial_t \Pi &=& - N \, \gamma^{ij} \, {\cal D}_i \Phi_j 
+ \beta^i {\cal D}_i \Pi \nonumber \\
&& - \frac{\Phi_i {\cal D}_j (N \psi^6 \gamma^{ij})}{\psi^6} + \Pi NK, 
\label{e:dtPi}\\
	\partial_t \Phi_i &=& - {\cal D}_i (N \, \Pi) + \Phi_k {\cal D}_i \beta^k 
+ \beta^k {\cal D}_k \Phi_i.
\eea
This system is solved using a second-order Adams-Bashforth scheme, with the 
time step being a submultiple of the one used for the evolution of the boundary 
conditions for the metric. Note that, in the case of the maximal slicing condition 
$(K=0)$, the last term of Eq.~(\ref{e:dtPi}) vanishes.

Matching across different domains is done along characteristic fields, in an 
upwind manner. At the outer boundary ($r=R_{\rm max}$), before the compactified
domain, we impose a Sommerfeld-like condition,
\begin{equation}
  \label{e:BC_pi_phir}
  \partial_t \left( \Pi - \Phi_r \right)_{r=R_{\rm max}}  = 0,
\end{equation}
and the consistency condition for $\phi$,
\begin{equation}
  \label{e:BC_phi}
  \left. \partial_t \phi \right|_{r=R_{\rm max}} = \left( -N\Pi +
    \beta^r \Phi_r \right)_{r=R_{\rm max}} .
\end{equation}
No boundary condition is needed for either field at the inner boundary 
(excision at $r=0.916\ M_{\rm ADM}$),as all characteristics are directed out of
the computational domain as long as the excision surface is spacelike, which is
verified in our case with $b-N>0$ in our numerical simulations. All matching 
and boundary conditions are implemented in a collocation approach to spectral 
variables. 

\section{Perfect-fluid equations}
\label{app:perfect_fluid}
Einstein's equations, in the case of nonvacuum spacetimes, have to be solved 
coupled with the hydrodynamic equations for the evolution of matter which can 
be derived from the local conservation of baryon number and energy-momentum, 
respectively,
\be
	\nabla_\mu J^\mu = 0, \;\; \nabla_\mu T^{\mu\nu} = 0,
\label{e:hydro}
\ee
with the current $J^\mu$ and the energy momentum-tensor $T^{\mu\nu}$ of a perfect 
fluid being
\be
	J^\mu = \rho \, u^\mu, \;\; 
T^{\mu\nu} = \rho \, h\, u^\mu \, u^\nu + p \, g^{\mu\nu},
\ee
where $\rho$ is the rest-mass (baryon mass) density, $u^\mu$ is the 4-velocity 
of the fluid, $h = 1+ \epsilon + p/\rho$ is the specific enthalpy, $\epsilon$ 
is the specific internal energy and $p$ is the pressure. The previous system of 
equations~(\ref{e:hydro}) can be written as a first-order hyperbolic system for
the conserved variables
$(D, S^j, \tau) = (\rho\, W, \rho\, h\, W\, v^j, \rho\, h\, W^2 -p- \rho\, W)$,
with $W$ being the Lorentz factor \cite{font-08}.


\bibliographystyle{apsrev}


\end{document}